\documentclass{article}
\usepackage{graphicx}%style pour inserer figures en eps
\usepackage{here}%pour inserer figures et tables a emplacement 
\def\beq{\begin{equation}}
\def\eeq{\end{equation}}
\def\bea{\begin{eqnarray}}
\def\eea{\end{eqnarray}}
\def\bq{\begin{quote}}
\def\eq{\end{quote}}

\def\nnb{\nonumber}
\def\ga{\left(}
\def\dr{\right)}

\def\rar{\rightarrow}
\def\lrar{\Longrightarrow}
\def\nnb{\nonumber}
\def\la{\langle}
\def\ra{\rangle}
\def\nin{\noindent}
\def\ba{\begin{array}}
\def\ea{\end{array}}

\def\b{\bullet}

\def\als{\alpha_s}
\def\as{\ga\frac{\bar{\alpha_s}}{\pi}\dr}

\begin{document}
\begin{flushright}
%PM/03-xx\\ 
\end{flushright}
\vspace*{5mm}
%\begin{center}
\section*{Open Charm and Beauty Chiral Multiplets in  QCD}
\vspace*{1.cm}
{\bf Stephan Narison}
\\
%\vspace{0.3cm}
Laboratoire de Physique Math\'ematique et Th\'{e}orique,
UM2, Place Eug\`ene Bataillon,
34095 Montpellier Cedex 05, France
\\ Email: narison@lpm.univ-montp2.fr
\\

\vspace*{.5cm}
%{\bf Abstract} \\ 
%\end{center}
%\vspace*{2mm}
\noindent
We study the dynamics of the the spin zero open charm and beauty mesons using QCD spectral sum rules (QSSR), where
we observe the important r\^ole of the chiral condensate $\la \bar \psi\psi\ra $ in the mass-splittings between the
scalar-pseudoscalar mesons. Fixing the sum rule parameters for reproducing the well-known $D(0^-)$ and $D_s(0^-)$ masses,
we re-obtain the running charm quark mass: $\bar m_c(m_c)=1.13^{+0.08}_{-0.04}$ GeV, which confirms our recent estimate from this
channel \cite{SNC}. Therefore, using sum rules {\it with no-free parameters}, we deduce $M_{D_s(0^+)}\simeq (2297\pm 113)$ MeV, which is
consistent with the observed $D_s(2317)$ meson, while a small $SU(3)$ breaking of about $25$ MeV
for the $D_s(0^+)-D(0^+)$ mass-difference has been obtained. We extend our analysis to the $B$-system and find $M_{B(0^+)}-
M_{B(0^-)}\simeq (422\pm 196)$ MeV confirming our old result from moment sum rules \cite{SNSP}. Assuming an approximate
(heavy and light) flavour and spin symmetries of the mass-splittings as indicated by our results, we also deduce
$M_{D^*_s(1^+)}\simeq (2440\pm 113)$ MeV in agreement with the observed
$D_{sJ}(2457)$.
We also get:
$f_{D(0^+)}\simeq (217\pm 25)$ MeV much bigger than $f_\pi$=130.6 MeV, while the size of the $SU(3)$ breaking
ratio $f_{D_s(0^+)}/f_{D(0^+)}\simeq 0.93\pm 0.02$ is opposite to the one of the $0^-$
channel of about 1.14. 
%%%%%%%%%%%%%%%%%%%%%%%
\section{Introduction}
The recent BABAR, BELLE and CLEO observations of two new states $D_{sJ}(2317)$ and $D_{sJ}(2457)$ \cite{BONDI} in the $D_s\pi$, $D_s\gamma$
and
$D_s\pi\gamma$ final states have stimulated a renewed interest in the spectroscopy of open
charm states which one can notice from different recent theoretical attempts to identify
their nature \cite{BONDI}. In this paper, we will try to provide the answer to this
question from QCD spectral sum rules \`a la Shifman-Vainshtein-Zakharov \cite{SVZ}. In fact, we have
already addressed a similar question in the past \cite{SNSP}, where we have predicted using QSSR the
mass splitting of the $0^+-0^-$ and $1^--1^+~\bar bu$ mesons using double ratio of moments sum rules
based on an expansion in the inverse of the $b$ quark mass. We found that the value of the
mass-splittings between the chiral multiplets were about the same and approximately independent of the spin
of these mesons signaling an heavy quark-type approximate symmetry:
\beq\label{eq: bsplit}
M_{B{(0^+)}}-M_{B{(0^-)}}\approx M_{B^*{(1^+)}}-M_{B^*{(1^-)}}\approx (417\pm 212)~{\rm MeV}~.
\eeq
The effect and errors on the mass-splittings are mainly due to the chiral condensate $\la\bar\psi\psi\ra$ and to the value of
the $b$ quark mass.
In this paper, we shall use an analogous approach to the open charm states. However, the same
method in terms of the $1/m_c$ expansion and some other nonrelativistic sum rules will be dangerous here
due to the relatively light value of the charm quark mass. Instead, we shall work with relativistic
exponential sum rules used successfully in the light quark channels for predicting the meson masses and QCD parameters \cite{SNB}
and in the
$D$ and
$B$ channels for predicting the (famous) decay constants $f_{D,B}$ \cite{SNC,SNB,SNFB} and the charm and bottom quark masses
\cite{SNC,SNB,SNFB,SNM,QMASS}.
\section{The QCD spectral sum rules}
 We shall work here with the (pseudo)scalar
two-point correlators: 
\bea
\psi_{P/S}(q^2) &\equiv& i \int d^4x ~e^{iqx} \
\la 0\vert {\cal T}
J_{P/S}(x)
J^\dagger _{P/S}(0) \vert 0 \ra ,
\eea
built from the (pseudo)scalar and (axial)-vector heavy-light quark currents:
\beq
J_{P/S}(x)=(m_Q\pm m_q)\bar Q(i\gamma_5)q,~~~~~ J^\mu_{V/A}=\bar Q\gamma^\mu(\gamma_5)q~.
\eeq
If we fix $Q\equiv c$ and $q\equiv s$, the corresponding mesons
have the quantum numbers of the $D_s(0^-)$, $D_s(0^+)$ mesons. $m_Q$ 
and $m_s$ are the running
quark masses. 
In the (pseudo)scalar channels, the relevant sum rules for our problem are the
Laplace transform sum rules:
\beq
{\cal L}^H_{P/S}(\tau)
= \int_{t_\leq}^{\infty} {dt}~\mbox{e}^{-t\tau}
~\frac{1}{\pi}~\mbox{Im} \psi^H_{P/S}(t),~~~{\mbox {and}}~~~
{\cal R}^H_{P/S}(\tau) \equiv -\frac{d}{d\tau} \log {{\cal L}^H_{P/S}(\tau)},
\eeq
where $t_\leq$ is the hadronic threshold, and H denotes the corresponding meson. The latter sum  rule,
 or its slight modification, is useful, as it is equal to the 
resonance mass squared, in  
 the simple duality ansatz parametrization of the spectral function:
\beq
\frac{1}{\pi}\mbox{ Im}\psi^H_P(t)\simeq f^2_DM_D^4\delta(t-M^2_D)
 \ + \ 
 ``\mbox{QCD continuum}" \Theta (t-t_c),
\eeq
where the ``QCD continuum comes from the discontinuity of the QCD
diagrams, which is expected to give a good smearing of the
different radial excitations \footnote{At
the optimization scale, its effect is negligible, such that a more
involved parametrization is not necessary.}. The decay constant $f_D$ is
analogous to $f_\pi=130.6$ MeV;
$t_c$ is the QCD continuum threshold, which is, like the 
sum rule variable $\tau$, an  (a priori) arbitrary 
parameter. In this
paper, we shall impose the
$\tau$- and $t_c$-stability criteria for extracting our optimal
results. The corresponding $t_c$ value also agrees with the FESR duality constraints \cite{RAF,SNB} and very roughly indicates
the position of the next radial excitations. However, in order to have a conservative result, we take a largest range of $t_c$ from the
beginning of $\tau$- to the one of $t_c$-stabilities. \\
The QCD expression of the correlator
is well-known to two-loop accuracy
(see e.g. \cite{SNB} and the explicit expressions given in \cite{SNC,SNFB}),
in terms  of the perturbative pole mass $M_Q$, and including the non-perturbative
condensates of dimensions less than or equal to six
\footnote{We shall
include the negligible contribution from the dimension six four-quark condensates, while we shall neglect an eventual
tachyonic gluon mass correction term found to be negligible in some other channels  \cite{ZAK}. }. For a
pedagocial presentation, we write the sum rule in the chiral limit ($m_s=0$), where the expression 
is more compact. In this way,  one can understand qualitatively the source of the mass splittings. The sum rule reads
to order $\alpha_s$:
\bea
{\cal L}^H_{P/S}(\tau)
&=& M^2_Q\Bigg{\{}\int_{M^2_Q}^{\infty} {dt}~\mbox{e}^{-t\tau}~\frac{1}{8\pi^2}\Bigg{[} 3 t(1-x)^2\Big{[}
1+\frac{4}{3}\as f(x)\Big{]}+{\cal O}(\alpha_s^2) \Bigg{]}\nnb\\
&&~\,\, +C_4\la O_4\ra_{P/S} +\tau C_6\la
O_6\ra_{P/S}~\mbox{e}^{-M^2_Q\tau}\Bigg{\}}~,
\eea
The different terms are \footnote{Notice a missprint in the expression given in \cite{SNSP,SNC}
which does not affect the result obtained there.}:
\bea
x&\equiv& M^2_Q/t,\nnb\\
f(x)&=&\frac{9}{4}+2\rm{Li}_2(x)+\log x \log (1-x)-\frac{3}{2}\log (1/x-1)\nnb\\
& & -\log (1-x)+ x\log (1/x-1)-(x/(1-x))\log x, \nnb\\
C_4\la O_4\ra_{P/S}&=&\mp M_Q\la \bar dd\ra~\mbox{e}^{-M^2_Q\tau} +\la \als G^2\ra\ga {3\over 2}-M_Q^2\tau\dr/12\pi\nnb\\
C_6\la O_6\ra_{P/S}&=&\mp\frac{M_Q}{2}\ga 1-\frac{M^2_Q\tau}{2}\dr
g\la\bar d\sigma_{\mu\nu}\frac{\lambda_a}{2}G_a^{\mu\nu}d\ra
\\ &&-\ga\frac{8\pi}{27}\dr\ga 2-\frac{M^2_Q\tau}{2}-\frac{M^4_Q\tau^2}{6}\dr\rho\als \la \bar
\psi\psi\ra^2~,
\eea
where we have used the contribution of the gluon condensate given in \cite{GEN}, which is IR finite
when letting $m_q\rar 0$ \footnote{The numerical change is negligible compared with the original expression obtained in \cite{NOVIKOV}.}.
The previous sum rules can be expressed in terms of the running mass $\bar{m}_Q(\nu)$
%\footnote{It is clear that, for the non-perturbative terms which are known to leading order
%of perturbation theory, one can use either the running or the pole mass. However~, we shall see
%that this distinction
%does not affect notably the present result.},
 through the perturbative  two-loop relation \cite{TARRACH}:
\bea\label{relation}
M_{Q}&=&\bar m_Q(p^2)\Bigg{[}1+\ga\frac{4}{3}+\ln{\frac{p^2}{M_Q^2}}\dr\as
+{\cal O}(\alpha_s^2)\Bigg{]}~,
\eea 
where $M_Q$ is the pole mass. 
Throughout this paper we
shall use the values of the QCD parameters given in Table 1. 
\begin{table}[H]
\begin{center}
% space before first and after last column: 1.5pc
% space between columns: 3.0pc (twice the above)
\setlength{\tabcolsep}{.28pc}
% -----------------------------------------------------
% adapted from TeX book, p. 241
\newlength{\digitwidth} \settowidth{\digitwidth}{\rm 1.5}
%\catcode`?=\active \def?{\kern\digitwidth}
% -----------------------------------------------------
\caption{QCD input parameters used in the analysis.}
%\begin{tabular*}{\textwidth}{@{}l@{\extracolsep{\fill}}rrrrr}
\begin{tabular}{ll}
\\
\\
\hline 
%\hline
  \\
Parameters& References\\
\\
\hline 
\\
$\Lambda_4=(325\pm  43)$ MeV&\cite{SNB}\\
$\Lambda_5=(225\pm 30)$ MeV&\cite{SNB}\\
$\bar m_b(m_b)=(4.24\pm 0.06)$ GeV&\cite{SNB,QMASS,SNC,HEAVYQUARK}\\
$\bar m_s(2~{\rm GeV})= (111\pm 22)$ MeV&\cite{SNB,QMASS,SNMS,JAMINA}\\
%$\bar m_c(m_c)=(1.23\pm 0.05)$ GeV&\cite{SNB,QMASS,SNC}\\
$\la \bar dd\ra^{1/3}$(2 GeV)=$-(243\pm 14)$ MeV&\cite{SNB,QMASS,DOSCHSN}\\
$\la \bar ss\ra /\la \bar dd\ra=0.8\pm 0.1$&\cite{SNB,SNP2}\\
$\la \alpha_s G^2\ra=(0.07\pm 0.01)$ GeV$^4$&\cite{SNB,SNG}\\
$M^2_0=(0.8\pm 0.1)$ GeV$^2$&\cite{SNB,SNSP}\\
$\alpha_s\la\bar\psi\psi\ra^2=(5.8\pm 2.4)\times 10^{-4}~$GeV$^6$&\cite{SNB,SNG,LNT}\\
%&\\
\\
\hline 
%\hline
\end{tabular}
\end{center}
\end{table}
\nin
We have used for the mixed condensate the
parametrization:
\bea
g\la\bar d\sigma_{\mu\nu}\frac{\lambda_a}{2}G_a^{\mu\nu}d\ra&=&M^2_0\la\bar dd\ra,
\eea
and deduced the value of the QCD scale $\Lambda$ from the value of $\alpha_s(M_Z)=(0.1184\pm 0.031)$
\cite{PDG,BETHKE}. We have taken the mean value of $m_s$ from recent reviews \cite{SNB,QMASS,JAMINA}.
%%%%%%%%%%%%%%%%%%%%%%%%%%%%%%%%%%
\section{Calibration of the sum rule from the $D(0^-)$ and $D_s(0^-)$ masses
and re-estimate of $\bar m_c(m_c)$}
\begin{figure*}[hbt]
\begin{center}
\includegraphics[width=7cm]{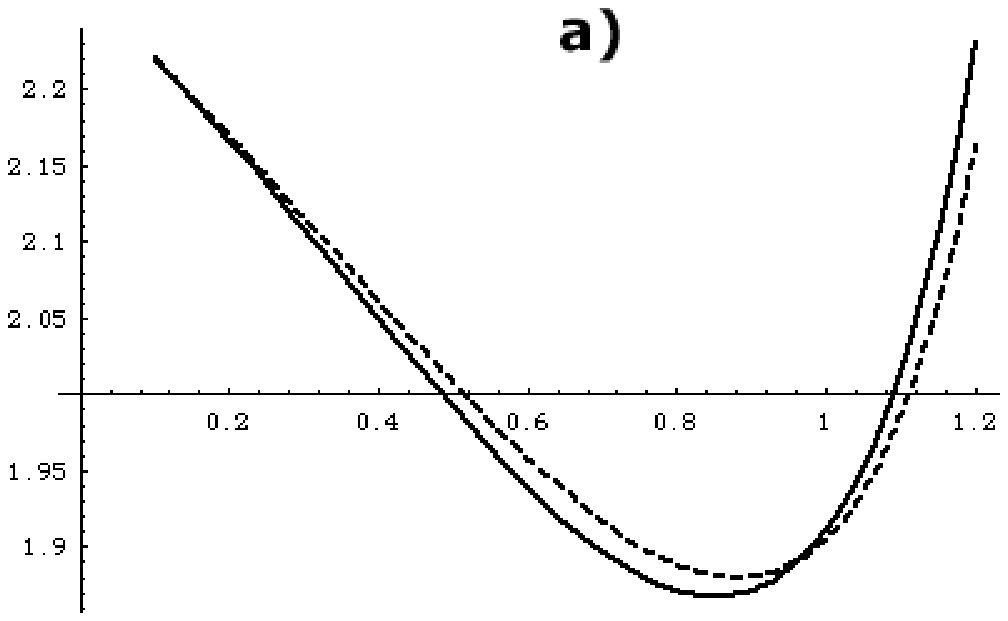}
\includegraphics[width=6cm]{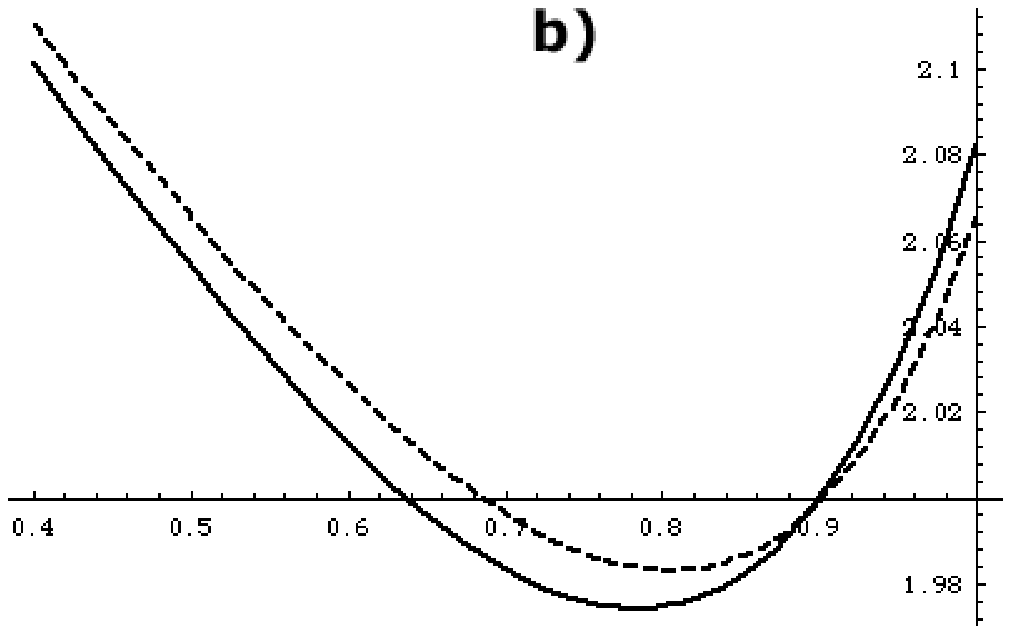}
\caption{$\tau$ in GeV$^{-2}$-dependence of the a) $M_D(0^-)$ in GeV for $\bar m_c(m_c)=1.11$ GeV and b) $M_{D_s(0^-)}$ in GeV for $\bar
m_c(m_c)=1.15$ GeV at a given value of $t_c=7.5$ GeV$^2$. The dashed line is the result including the leading
$\la \bar \psi\psi\ra$ contribution. The full line is the one including non-perturbative effects up to dimension-six.} 
\end{center}
\end{figure*}
\nin
This analysis has been already done in previous papers to order $\alpha_s$ and $\alpha_s^2$ and has served to fix the running charm quark
mass. We repeat this analysis here to order $\alpha_s$ for a pedagogical purpose. We show in Fig. 1a), the $\tau$-dependence of the $D(0^-)$
and in Fig 1b) the one of the
$D_s(0^-)$ masses for a given value of $t_c$, which is the central value of the range:
\beq
t_c=(7.5\pm 1.5)~{\rm GeV}^2~,
\eeq
where the lowest value corresponds to  the beginning of $\tau$-stablity  and the highest one to the beginning of $t_c$ stability obtained by 
\cite{SNC,SNFB,SNB} in the analysis of $f_D$ and $f_{D_s}$. This range of $t_c$-values covers the different choices of $t_c$ used in the sum
rule literature. As mentioned previously, the one of the beginning of $t_c$-stability co\"\i ncides, in general, with the value obtained
from FESR local duality constraints \cite{RAF,SNB}. Using the input values of QCD parameters in Table 1,  the best fits of
$D(0^-)$ (resp.
$D_s(0^-))$ masses for a given value of
$t_c=7.5$ GeV$^2$ correspond to a value of $\bar m_c(m_c)$ of 1.11 (resp. 1.15) GeV. Taking the mean value as an estimate, one can deduce:
\beq\label{cmass}
\bar m_c(m_c^2)=(1.13^{+0.07}_{- 0.02}\pm 0.02\pm 0.02\pm 0.02)~{\rm GeV}~,
\eeq
where the errors come respectively from $t_c$, $\la\bar\psi\psi\ra$, $\Lambda$ and the mean value of $m_c$
required from fitting the 
$D(0^-)$ and $D_s(0^-)$ masses. 
This value is perfectly consistent with the one obtained in  \cite{SNC,SNFB} obtained to the same order and to order $\alpha_s^2$,
indicating that, though the $\alpha_s^2$ corrections are both large in the two-point function and $m_c$ \cite{CHET2}, it does not affect
much the final result from the sum rule analysis. In fact, higher corrections tend mainly to shift the
position of the stability regions but affect slightly the output value of $m_c$. This value of
$m_c$ is in the range of the current average value
$(1.23\pm 0.05)$ GeV reviewed in \cite{SNB,QMASS,PDG}. However, it does not favour higher values of $m_c$ allowed in some other channels and
by some non relativistic sum rules and approaches. However, these non relativistic approaches might be quite inaccurate due to
the relative smallness of the charm quark mass. Higher values of $m_c$ would lead to an overestimate of the
$D(0^-)$ and $D_s(0^-)$ masses.  In the following analysis, we shall use the central value $\bar m_c(m_c)=1.11$ (resp. 1.15) GeV for the
non-strange (resp. strange) meson channels.
%%%%%%%%%%%%%%%%%%%%%%%%%%%%%%%%%%%%%%%%%%%
\section{The $0^+-0^-$ meson mass-splittings}
\begin{figure}[hbt]
\begin{center}
\includegraphics[width=7cm]{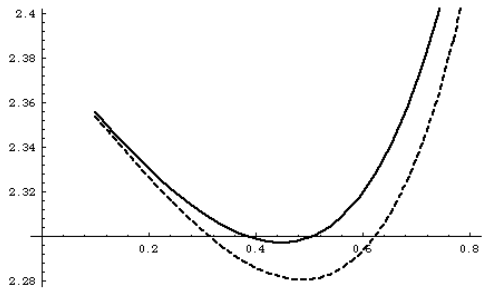}
\caption{Similar to Fig. 1 but $\tau$-behaviour of $M_{D_s(0^+)}$ for given values of  $t_c=7.5$ GeV$^2$ and $m_c(m_c)=1.15$ GeV.} 
\end{center}
\end{figure}
\nin
\begin{itemize}
\item We study in Fig. 2), the $\tau$-dependence of the $D_s(0^+)$ mass at the values of $t_c$ and $m_c$ obtained
previously.  In this way, we obtain:
\beq\label{eq: dssp}
M_{D_s{(0^+)}}\simeq (2297^{+81+63}_{- 98-70}\pm 11\pm 2\pm 11)~{\rm MeV}~~~~~\lrar~~~~~ M_{D_s{(0^+)}}-M_{D_s{(0^-)}} = (328\pm 113)~{\rm MeV}~,
\eeq
where the errors come respectively from $t_c$, $m_c$, $\la\bar\psi\psi\ra$, $m_s$, and $\Lambda$. 
We have used the experimental value of $M_{D_s{(0^-)}}$.
 The reduction of the theoretical error needs
precise values of the continuum threshold  \footnote{The range of $t_c$ values 6-9 GeV$^2$ obtained previously for the $D(0^-)$ mesons
co\"\i ncides a posteriori with the corresponding range for the $D(0^+)$ meson if one assumes that the splitting between the radial
excitations is the same as the one between the ground states, i.e about 300 MeV. We have cheked during the analysis that this effect is
unimportant and is inside the large error induced by the range of $t_c$ used.} and of the charm quark mass which are not within the present
reach of the estimate of these quantities. Further discoveries of the continuum states will reduce the present error in the splitting. One
should also notice that in the ratio of sum rules with which we are working, we expect that perturbative radiative corrections are
minimized though individually large in the expression of the correlator and of the quark mass.
\item 
The value of the mass-splittings obtained previously is comparable with the one of the $B(0^+)$-$B(0^-)$ given in Eq. (\ref{eq:
bsplit}), and suggests an approximate heavy-flavour symmetry of this observable.
\item
We also derive the result in the limit of $SU(3)_F$ symmetry where the strange quark
mass is put to zero, and where the $\la\bar\psi\psi\ra$ condensate is chirally symmetric ($\la\bar
ss\ra=\la\bar dd\ra$). In this case, one can predict an approximate degenerate mass within the errors:
\beq\label{eq: su3split}
M_{D_s{(0^+)}}-M_{D{(0^+)}}\simeq 25 ~{\rm MeV}~,
\eeq
which indicates that the mass-splitting between the strange and non-strange $0^+$ 
open charm mesons is almost not affected by $SU(3)$ breakings, contrary to the case of the $0^-$ mesons with a splitting of about 100 MeV. 
\item We extend the analysis to the case of the $B(0^+)$ meson. Here, it is more informative to predict the ratio of the $0^+$ over the $0^-$
masses as the prediction on the absolute values though presenting stability in $\tau$ tend to overestimate the value of $M_B$. We obtain:
\beq\label{eq: bsplit2}
{M_{B{(0^+)}}\over M_{B{(0^-)}}}\simeq (1.08\pm 0.03\pm 0.03\pm 0.02\pm 0.02)~~~~~~
\lrar~~~~~~M_{B{(0^+)}}-M_{B{(0^-)}}\simeq (422\pm 196) ~{\rm MeV}~, 
\eeq
where the errors come respectively from $t_c$ taken in the range $43-60$ GeV$^2$, $m_b$, $\la\bar\psi\psi\ra$, and $\tau$. It
agrees with the result in Eq. (\ref{eq: bsplit}) obtained from moment sum rules \cite{SNSP}. We have used the value of $\bar
m_b(m_b)$  given in Table 1.
\end{itemize}
%%%%%%%%%%%%%%%%%%%%%%%%%%%%%%%%%%%%%%%
\section{The $1^+-1^-$ meson mass-splitting}
Our previous results in Eqs. (\ref{eq: bsplit}), (\ref{eq: dssp}) to (\ref{eq: bsplit2}) suggest that the mass-splittings are approximately
 (heavy and light) flavour and spin independent. Therefore, one can write to a good approximation the empirical relation:
\beq\label{eq: emp}
M_{D_s{(0^+)}}-M_{D_s{(0^-)}}\approx M_{D{(0^+)}}-M_{D{(0^-)}}\approx M_{B{(0^+)}}-M_{B{(0^-)}}\approx 
M_{B{(1^+)}}-M_{B{(1^-)}}\approx M_{D^*_s{(1^+)}}-M_{D^*_s{(1^-)}}~.
\eeq
Using the most precise number given in Eq. (\ref{eq: dssp}), one can deduce:
\beq\label{eq: 1split}
M_{D^*_s{(1^+)}}-M_{D^*_s{(1^-)}}\simeq  (328\pm 113)~{\rm
MeV}~~~~~\lrar~~~~~M_{D^*_s{(1^+)}}\simeq (2440\pm 113)~{\rm MeV}~.
\eeq
This result is consistent with the $1^+$ assignement of the $\bar cs$ meson $D_{sJ}(2457)$ discovered recently \cite{BONDI}. In a future work, we
plan to study in details this spin one channel using QSSR.

%%%%%%%%%%%%%%%%%%%%%%%%%%%%%%%%%%%%%%%%%%%%%%%%
\section{The $D(0^+)$ and $D_s(0^+)$  decay constants}
\begin{figure}[hbt]
\begin{center}
\includegraphics[width=7cm]{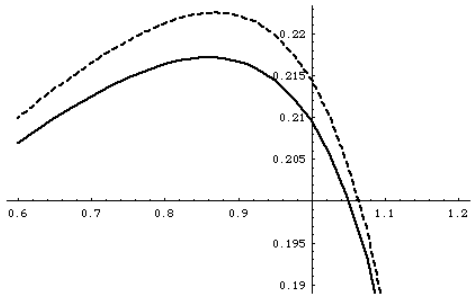}
\caption{ Similar to Fig. 1 but $\tau$-behaviour of $f_D(0^+)$ for given values of  $t_c=7.5$ GeV$^2$ and $\bar m_c(m_c)=1.11$ GeV.} 
\end{center}
\end{figure}
\nin
For completing our analysis, we estimate the decay constant $f_{D(0^+)}$ analogue to $f_\pi=130.6$ MeV.
We show the behaviour of $f_{D(0^+)}$ versus $\tau$, where a goood stablity is
obtained. Adopting the range of $t_c$-values obtained previously and using $\bar m_c(m_c)=1.11^{+0.08}_{-0.04}$ GeV required for a best fit of
the non strange $D(0^+)$ meson mass,
we deduce to two-loop accuracy:
\beq\label{eq: fdplus}
f_{D(0^+)}=(217^{+5+15}_{-15-19} \pm 10\pm 10)~{\rm MeV}~,
\eeq
where the errors come respectively from the values of $t_c$, $m_c$, $\la\bar\psi\psi\ra$ and $\Lambda$. We have
fixed
$M_{D(0+)}$ to be about 2272 MeV from our previous fit. It is informative to compare this result with the one of $f_{D}=(205\pm
20)$ MeV, where the main difference can be attributed by the sign flip of the quark condensate contribution in the QCD expression
of the corresponding correlators. A numerical study of the $SU(3)$ breaking effect leads to:
\beq\label{decay1}
r_s\equiv\frac{f_{D_s(0^+)}}{f_{D(0^+)}}\simeq 0.93\pm 0.02~,
\eeq
which is reverse to the analogous ratio in the pseudoscalar channel $f_{D_s}/f_D\simeq 1.14\pm 0.04$ given semi-analytically in
\cite{SNFBS}. In order to understand this result, we give a semi-analytic parametrization of this $SU(3)$ breaking ratio. Keeping
the leading term in $m_s$ and $\la\bar\psi\psi\ra$, one obtains:
\beq\label{decay2}
r_s\simeq \ga 1-{m_s\over m_c}\dr\Big{[} 1-7.5\la \bar ss-\bar dd\ra\Big{]}^{1/2}\ga{M_{D_s(0^+)}\over M_{D(0^+)}}\dr^2\simeq 0.9~,
\eeq
where the main effect comes from the negative sign of the $m_s$ contribution in the overall normalization of the scalar current,
while the meson mass ratio does not compensate this effect because of the almost equal mass of $D_s(0^+)$ and $D(0^+)$ obtained in
previous analysis. This feature is opposite to the case of $f_{D(0^-)}$.
%%%%%%%%%%%%%%%%%%%%%%%%%%%%%%%%%%
\section{Summary and conclusions}
Due to the experimental recent discovery of the $D_{sJ}(2317)$ and $D^*_{sJ}(2457)$, we have analyzed using QSSR the dynamics of the $0^\pm$
and
$1^\pm$ open charm and beauty meson channels: 
\begin{itemize}
\item We have re-estimated the running charm quark mass from the $D$ and $D_s$ mesons. The result in Eq. (\ref{cmass}) confirms earlier
results obtained to two- and three-loop accuracies \cite{SNFB,SNC}.
\item We have studied the mass-splittings of the
$0^+$-$0^-$ in the
$D$ systems using QSSR. Our result in the $(0^+)$ channel given in Eq. (\ref{eq: dssp}) agrees with the
recent experimental findings of the $D_{sJ}(2317)$ suggesting that this state is a good candidate for being a $\bar cs$ $0^+$ meson.
\item We also found, in Eq. (\ref{eq: su3split}), that the $SU(3)$ breaking responsible of the mass-splitting between the $D_s(0^+)$ and
$D(0^-)$ is small of about 25 MeV
 contrary to the case of the pseudoscalar $D_s$-$D$ mesons of about 100 MeV. 
\item We have
extended our analysis to the $B$-system. Our results in Eqs. (\ref{eq: dssp}), (\ref{eq: bsplit2}) and (\ref{eq: bsplit}) suggest an
approximate (light and heavy) flavour and spin symmetries of the meson mass-splittings. We use this result to get the mass of the $\bar
cs~D^*_s(1^+)$ meson in Eq. (\ref{eq: 1split}), which is in (surprising) good agreement with the observed
$D^*_{sJ}(2457)$. 
\item Using QSSR, we have also determined the decay constants of the $0^+$ mesons and compare them
with the ones of the
$0^-$ states. The result in Eq. (\ref{eq: fdplus}), which is similar to
the pseudoscalar decay constant $f_D\simeq 205$ MeV, suggests a huge violation of the heavy quark symmetry $1/\sqrt{M_D}$ scaling law.
Finally, our results in Eqs. (\ref{decay1}) and (\ref{decay2}) indicate that the $SU(3)$ breaking act in an opposite way compared to the case
of the $0^-$ channels.
\end{itemize}
Experimental or/and lattice measurements of the previous predictions are useful for testing the validity of the results
obtained to two-loop accuracy in this paper from QCD spectral sum rules. However, a complete confirmation of the nature of these new states
needs a detail study of their production and decays. We plan to come back to these points in a future work.\\
\\
During the editorial preparation of this paper, there appears in the literature a paper using sum rules method to
the same channel [25]. Instead of our result in Eq. (13): $M_{D_s(0+)}=(2297\pm 113)$ MeV, the authors obtain
$M_{D_s(0+)}=(2480\pm 30)$ MeV which is higher than the BABAR result $D_s(2317)$ by about 160 MeV. Considering
this deviation as significant, the authors conclude that the experimental candidate cannot be a $\bar cs$ state
contrary to the conclusion reached in the present paper. However, by scrutinizing the analysis of Ref. [25], we find that the true
errors of the analysis have been underestimated:\\
$\b$ As one can see from their
figures, the quoted error of 30 MeV only takes into account the one due to
the choice of the QCD contiuum threshold taken in a smaller range 8.1-9.3 GeV$^2$, than the more
conservative value 6-9 GeV$^2$ used in the present paper, which induces a larger error of 80-98 MeV. \\
$\b$ The high central value obtained in [25] is related to a higher choice of the (ill-defined) charm quark pole
mass of 1.46 GeV compared to the value 1.3 GeV which would have been deduced from its relation with the running charm quark mass 1.15
GeV in Eq. (9) used in this paper to reproduce the $D_s(0^-)$ mass. As the analysis is very sensitive to $m_c$, its induced error should be
included in the final number. With $m_c$ in Eq. (12), this uncertainty is about 70 MeV, and might be bigger if one considers
all range of $m_c$ given in the literature. In this paper, we have used $M_D$ and$M_{D_s}$ as alternative channels for extracting $m_c$.\\
$\b$ Taking properly the different sources of errors including the running of condensate and the effects of some
other anomalous dimensions, one then leads to a consistency within the errors of both theoretical estimates with the
present data.

%\vfill\eject
%\input{bibi}
%

\end{document}